\documentclass[12pt]{article}

\usepackage{epsfig,epsf}

\begin{document}

\def\beq{\begin{equation}}

\def\eeq{\end{equation}}
\def\eq#1{{Eq.~(\ref{#1})}}

\def\fig#1{{Fig.~\ref{#1}}}

\newcommand{\bas}{\bar{\alpha}_S}

\newcommand{\as}{\alpha_S} 

\newcommand{\bra}[1]{\langle #1 |}

\newcommand{\ket}[1]{|#1\rangle}

\newcommand{\bracket}[2]{\langle #1|#2\rangle}

\newcommand{\intp}[1]{\int \frac{d^4 #1}{(2\pi)^4}}

\newcommand{\mn}{{\mu\nu}}

\newcommand{\tr}{{\rm tr}}

\newcommand{\Tr}{{\rm Tr}}

\newcommand{\T} {\mbox{T}}

\newcommand{\braket}[2]{\langle #1|#2\rangle}

\newcommand{\ab}{\bar{\alpha}_S}

\setcounter{secnumdepth}{7}

\setcounter{tocdepth}{7}

\parskip=\itemsep               

\setlength{\itemsep}{0pt}       

\setlength{\partopsep}{0pt}     

\setlength{\topsep}{0pt}        


\setlength{\textheight}{22cm}

\setlength{\textwidth}{174mm}

\setlength{\topmargin}{-1.5cm}




\newcommand{\beqar}[1]{\begin{eqnarray}\label{#1}}

\newcommand{\eeqar}{\end{eqnarray}}

\newcommand{\m}{\marginpar{*}}

\newcommand{\lash}[1]{\not\! #1 \,}

\newcommand{\nn}{\nonumber}

\newcommand{\D}{\partial}

\newcommand{\h}{\frac{1}{2}}

\newcommand{\g}{{\rm g}}

\newcommand{\el}{{\cal L}}

\newcommand{\A}{{\cal A}}

\newcommand{\Ka}{{\cal K}}

\newcommand{\al}{\alpha}

\newcommand{\be}{\beta}

\newcommand{\ep}{\varepsilon}

\newcommand{\ga}{\gamma}

\newcommand{\de}{\delta}

\newcommand{\De}{\Delta}

\newcommand{\et}{\eta}

\newcommand{\ka}{\vec{\kappa}}

\newcommand{\la}{\lambda}

\newcommand{\ph}{\varphi}

\newcommand{\si}{\sigma}

\newcommand{\ro}{\varrho}

\newcommand{\Ga}{\Gamma} 

\newcommand{\om}{\omega}

\newcommand{\La}{\Lambda}  

\newcommand{\tG}{\tilde{G}}

\renewcommand{\theequation}{\thesection.\arabic{equation}}



%

\def\ap#1#2#3{     {\it Ann. Phys. (NY) }{\bf #1} (19#2) #3}

\def\arnps#1#2#3{  {\it Ann. Rev. Nucl. Part. Sci. }{\bf #1} (19#2) #3}

\def\npb#1#2#3{    {\it Nucl. Phys. }{\bf B#1} (19#2) #3}

\def\plb#1#2#3{    {\it Phys. Lett. }{\bf B#1} (19#2) #3}

\def\prd#1#2#3{    {\it Phys. Rev. }{\bf D#1} (19#2) #3}

\def\prep#1#2#3{   {\it Phys. Rep. }{\bf #1} (19#2) #3}

\def\prl#1#2#3{    {\it Phys. Rev. Lett. }{\bf #1} (19#2) #3}

\def\ptp#1#2#3{    {\it Prog. Theor. Phys. }{\bf #1} (19#2) #3}

\def\rmp#1#2#3{    {\it Rev. Mod. Phys. }{\bf #1} (19#2) #3}

\def\zpc#1#2#3{    {\it Z. Phys. }{\bf C#1} (19#2) #3}

\def\mpla#1#2#3{   {\it Mod. Phys. Lett. }{\bf A#1} (19#2) #3}

\def\nc#1#2#3{     {\it Nuovo Cim. }{\bf #1} (19#2) #3}

\def\yf#1#2#3{     {\it Yad. Fiz. }{\bf #1} (19#2) #3}

\def\sjnp#1#2#3{   {\it Sov. J. Nucl. Phys. }{\bf #1} (19#2) #3}

\def\jetp#1#2#3{   {\it Sov. Phys. }{JETP }{\bf #1} (19#2) #3}

\def\jetpl#1#2#3{  {\it JETP Lett. }{\bf #1} (19#2) #3}


\def\ppsjnp#1#2#3{ {\it (Sov. J. Nucl. Phys. }{\bf #1} (19#2) #3}

\def\ppjetp#1#2#3{ {\it (Sov. Phys. JETP }{\bf #1} (19#2) #3}

\def\ppjetpl#1#2#3{{\it (JETP Lett. }{\bf #1} (19#2) #3} 

\def\zetf#1#2#3{   {\it Zh. ETF }{\bf #1}(19#2) #3}

\def\cmp#1#2#3{    {\it Comm. Math. Phys. }{\bf #1} (19#2) #3}

\def\cpc#1#2#3{    {\it Comp. Phys. Commun. }{\bf #1} (19#2) #3}

\def\dis#1#2{      {\it Dissertation, }{\sf #1 } 19#2}

\def\dip#1#2#3{    {\it Diplomarbeit, }{\sf #1 #2} 19#3 }

\def\ib#1#2#3{     {\it ibid. }{\bf #1} (19#2) #3}

\def\jpg#1#2#3{        {\it J. Phys}. {\bf G#1}#2#3}  

%


%


\def\thefootnote{\fnsymbol{footnote}} 

%

%

%


\noindent

\begin{flushright}

\parbox[t]{10em}{

 \today \\
DESY-03-116
}
\end{flushright}

\vspace{1cm}

\begin{center}

{\LARGE  \bf   A Linear Evolution for Non-Linear Dynamics  
and Correlations in Realistic Nuclei}\\

\vskip1cm

{\large \bf ~E. ~Levin ${}^{a),b) \,\ddagger}$ 
\footnotetext{${}^{\ddagger}$ \,\,Email:
leving@post.tau.ac.il, levin@mail.desy.de.}  
and ~M. ~Lublinsky ${}^{b) \,\star}$ 

\footnotetext{${}^{\,\star}$ \,\,Email:
lublinm@mail.desy.de }}

\vskip1cm

{\it ${}^{a)}$\,\,\, HEP Department}\\
{\it School of Physics and Astronomy}\\
{\it Raymond and Beverly Sackler Faculty of Exact Science}\\
{\it Tel Aviv University, Tel Aviv, 69978, Israel}\\
\vskip0.3cm
{\it ${}^{b)}$  DESY Theory Group, DESY}\\
{\it D-22607 Hamburg, Germany}\\
\vskip0.3cm

\end{center}  

\bigskip

\begin{abstract}        
A new approach to high energy evolution based on a linear
equation for QCD generating functional is developed. This
approach opens a possibility for systematic study of 
correlations inside targets, and, in particular, inside realistic 
nuclei.

Our results are presented as three new equations.
The first one is a linear equation  for QCD generating functional 
(and for scattering amplitude) that sums the 'fan' 
diagrams. For the amplitude this equation is equivalent to the non-linear 
Balitsky-Kovchegov equation.

The second equation is a generalization of the 
Balitsky-Kovchegov non-linear  equation to interactions with  
realistic nuclei. It includes a new correlation parameter which 
incorporates, in a model dependent way, correlations inside the nuclei. 

The third equation is a non - linear  equation for QCD generating 
functional (and for scattering amplitude) that in addition to 
the 'fan' diagrams sums  the Glauber-Mueller multiple rescatterings.  
\end{abstract}



\newpage

\def\thefootnote{\arabic{footnote}} 

\section{Introduction}

\label{sec:Introduction}

High density QCD \cite{GLR,MUQI,MV,SAT,ELTHEORY} 
which is a theory of Color Glass Condensate
deals with  parton systems with large  gluon 
occupation numbers. It  has entered  a new phase of its 
development: a direct comparison with the experimental data.  A 
considerable success \cite{GW,KS,Munier,GLM,LUB,BGLLM,ESK,KWT,KML,KLN} 
has been reached in 
description of new precise 
data 
on deep inelastic scattering \cite{HERADATA} as well as in understanding 
of general features of hadron production in ion-ion collision 
\cite{RHICDATA}.

Most of the applications of hdQCD
are based on or related to a nonlinear evolution
equation derived for high density QCD \cite{GLR,MUQI,B,K}. The equation
sums 'fan' diagrams, which is a subset of the semi-enhanced diagrams.
In order to obtain a closed form equation it is usually  assumed that
color dipoles produced  by the evolution interact independently \cite{GLR,K}.
It was noticed in 
Ref. \cite{BGLLM} that the nonlinear evolution equation based on the above 
assumption faces problems when applied to realistic nuclei. 
It was realized that in addition
to multiple rescatterings on nucleons inside a nucleus we have to include
in the analysis  multiple rescatterings inside a single nucleon. This
means a need to account for nucleus correlations. It is our motivation
for this research to develop  a systematic approach which would allow
us to introduce correlations inside a target. 

We have found that a convenient language to address the problem of
target correlations is using the QCD generating functional, 
for which a linear evolution equation can be written. The generating
functional was introduced by Mueller in Ref. \cite{MUUN} and then
used by Kovchegov in his derivation of the non-linear evolution equation
for interaction amplitude \cite{K}. In the present paper, we show that
this functional (and the amplitude) obeys a linear evolution equation
involving functional derivatives with respect to initial conditions.  
For the linear equation we propose a systematic method to include target
correlations. In general, the linear equation for the amplitude with
target correlations cannot be transformed to a non-linear one. For a 
specific choice of correlations we are able  to reduce the linear equation
to a non-linear form similar to the Balitsky-Kovchegov (BK) equation. 
In fact, our new equation
is a model dependent generalization of the BK equation for targets with 
correlations. It is important to stress that we consider the very same
functional as of Mueller. The evolution of the dipole wave function is
still based on independent production of dipoles. Only correlations 
for the dipole-target interaction are introduced. 

Let us discuss now realistic nuclei and reasons why we believe target
correlations are essential there.
In principal, the BK 
equation is correct for hadron targets as well as for heavy nuclei.
For the latter the large $A$ limit is usually assumed which  adds an 
additional justification to the equation validity. In this case  a nucleus
is viewed as a very dense system of nucleons. We are going to 
relax  this assumption. In fact the realistic nuclei are not very dense but
rather dilute \cite{LUB,KWT}. 
The diluteness parameter that governs the interaction with nuclei 
could be expressed in the form:
\beq \label{KAPPA}
\kappa_A(b)\,\,=\,\,\pi \,R^2_N\,\,S_A(b)\,,
\eeq
where $\pi R^2_N$ is the effective area in which  gluons of a nucleon are 
distributed. $S_A (b)$ stands for a nucleus profile function, which can be
taken as the Wood-Saxon  form factor of the nucleus.

The parameter $\kappa_A\,\propto\,\,A^{\frac{1}{3}}$ and , in 
principle, is large for heavy nuclei. It turns out, however, 
 that the gluons are distributed  inside a nucleon within 
a rather small area with radius of the order of $0.3-0.4 fm$ 
\cite{LUB,KWT}, which is much smaller compared to the proton radius. 
For such small area the value of $\kappa_A$ even for the heaviest nuclei 
is smaller than unity.  Therefore, in spite of the pure theoretical interest  
to the very heavy nuclei we have to be very careful applying the large $A$
limit to  the realistic nuclei. 
It should be stressed that even for  more `standard'  approach with the 
radius of gluon distribution of the order of $0.6 \,fm$ the value of 
$\kappa_A$ reaches at best 3 for  heaviest nuclei. In this case the 
$1/\kappa_A$ corrections   are of the order 35 \% and should be accounted
for when confronting with experimental data. We remind that
the most interesting effects in 
ion-ion and hadron-ion collisions such as Cronin enhancement  and 
$N_{part}$-scaling \cite{RHICDATA} are  of the level about 25\%. 

In this paper we  consider realistic nuclei without
the large $A$ assumption. We  argue that in this case a nonlinear evolution
is still a valid tool. However the equation gets modified compared to the one
of Ref. \cite{K} with which our discussion will have a strong overlap.
For realistic (dilute) nuclei the assumption about independent dipole 
interactions should be relaxed. Within some model we will be able to
include correlations inside nucleus such that only one correlation factor
is introduced. This factor is independent of the 
number of dipoles interacting with the nucleus.

The problem which we are going to deal with is very simply illustrated
on the following toy example\footnote{We are thankful to Alex Kovner
who understood our problem and brought in this example.}. 
Suppose we have a probe
with two dipoles only which are close to each other on the inter-nucleon
distance scale $T_A\,\sim A^{1/3}$. 
Suppose the effective interaction radius of a nucleon
is $R_N$. Then the probability for one dipole to interact with the nucleus
is $A\,R_N^2/T_A^2$. What will be the probability for two dipoles to interact
with the nucleus? For the case of independent scattering it is  
$(A\,R_N^2/T_A^2)^2$. It is obvious, however, that this is not what actually
happens. Since the dipoles are close to each other they either both hit the
nucleon or both miss it. The real probability will be rather  $A\,R_N^2/T_A^2$.
It follows from this discussion that the equation based on independent
interactions underestimates  shadowing corrections in nuclei. In the above 
example $\kappa_A\,=\,A\,R_N^2/T_A^2$.

If schematically the equation for the amplitude of Ref. \cite{K} has the form
\beq\label{KOV}
\frac{d N}{d Y}\,\,=\,\,Ker \otimes\,\left [\,N\,-\,N^2\right]\,
\eeq
then the equation which we propose for the dipole-nucleus amplitude can be
written as
\beq\label{we}
\frac{d N_A}{d Y}\,\,=\,\,Ker \otimes\,
\left[\,N_A\,-\,\frac{1\,+\,\kappa_A}{\kappa_A}\,N_A^2\right]\,
\eeq
\eq{we} can be certainly brought to the form of \eq{KOV} provided
a corresponding rescaling of the initial conditions is done. 
When $\kappa_A$ is large \eq{we} reduces to \eq{KOV}. In the opposite 
limit $\kappa_A$ is small, $N_A\,=\,\kappa_A\,N_N$ with $N_N$ being a solution
of \eq{KOV} for a nucleon target.

In the final part of this paper we generalize the approach based on QCD
generating functional to include Glauber-Mueller multiple rescatterings.
In order to achieve our goal we construct a QCD  effective 
 vertex for one dipole splitting to many. In Pomeron language, this vertex
generalizes the famous triple Pomeron vertex to the case of multiple splitting.
We define a new QCD generating functional which obeys a  new linear evolution 
equation. Without target correlations this equation can be reformulated
as a new non-linear equation. In addition to the 'fan' diagrams,
 the functional sums all types of semi-enhanced diagrams. We argue that in
the saturation domain all semi-enhanced diagrams are of equal importance. 
This is why they have to be summed up in order to get a reliable description
of the saturation domain. We find that the inclusion of the 
Glauber-Mueller multiple rescatterings leads to a significant modification
of results based on 'fan' diagrams (BK equation) only.

The structure of the paper is following. In the next section (2) we
derive a linear evolution equation for generating functional and the 
interaction amplitude including target correlations. Section 3 is devoted
to realistic nuclei. A model for nucleus correlations is proposed and
a generalization of the BK equation is given. In section 4 we present 
a generalization of the functional approach to Glauber-Mueller multiple
rescatterings. The concluding section 5 contains  discussion of the results
obtained.

\section{Linear Equation}

Though  we usually consider a nonlinear evolution for the scattering
amplitude it is possible to formulate the very same problem in terms of a
linear functional equation. A linear formulation appears to be more
suitable for treatment of correlations inside a target. 
We first start from a toy model which simplify the whole problem to 
such an extent that it could be solved analytically. Then we proceed
to the complete analysis of the QCD generating functional.

\subsection{A toy model}

In pQCD the Pomeron is described by the linear BFKL equation \cite{BFKL} which 
is integro-differential equation. Instead we will consider a phenomenological
Pomeron given by 
\beq \label{TM1}
\frac{d \,N(Y,\,b_t)}{d\, Y}\,\,=\,\,\omega_0\,N(Y,\,b_t)
\eeq
This equation reproduces the main property of the real BFKL equation,
namely, the  power-like increase of the amplitude as a function of energy. 
So far we neglect the fact that the actual degrees of freedom are
color  dipoles having different sizes. We will come back to the real QCD
in the next subsection.

\begin{figure}
\begin{center}
\epsfig{file=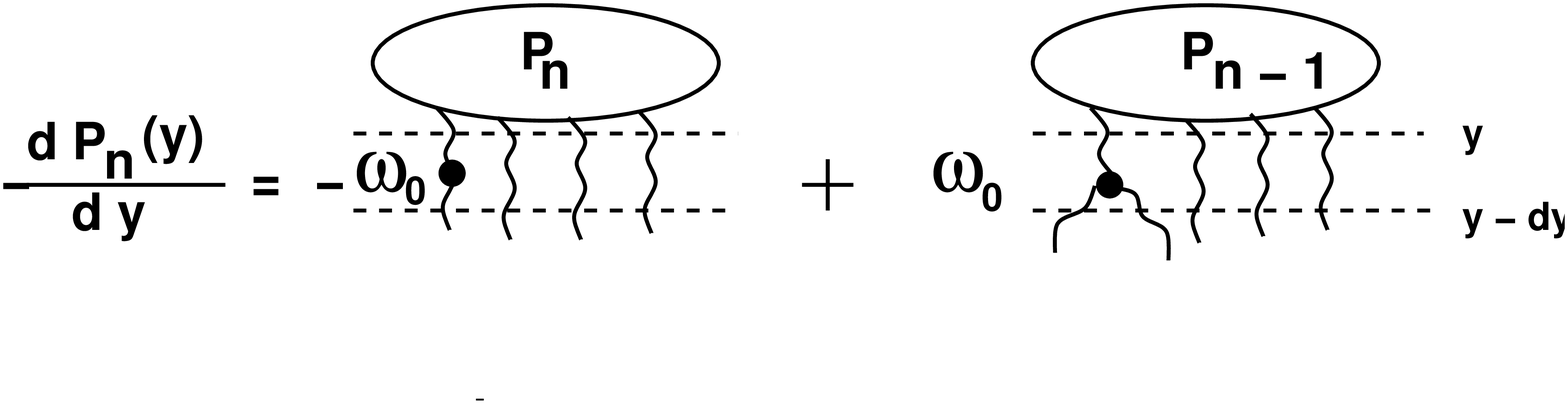,width=160mm}
\end{center}
\caption{The equation for the probability to find $n$-dipoles in one fast 
dipole. The wave line shows a single dipole.}
\label{pneq}
\end{figure}

Let $P_n(y)$ be a probability to find 
$n$-dipoles with rapidity $y$ in the wave function of the fastest (parent)
dipole moving with rapidity $Y\,>\,y$.
We  introduce in a very simple way a probability for
 one dipole to decay into two:
\beq \label{V} 
K(1\,\rightarrow 2)\,\,=\,\,\omega_0
\eeq 
In (\ref{V}) we have assumed that the triple Pomeron vertex is just 
$\omega_0\,\sim\,\alpha_s$.
For $P_n(y)$ we can easily write down a reccurent equation (see \fig{pneq} )
\beq \label{PNEQ}
-\,\frac{\partial\,P_n(y)}{\partial \,y}\,\,=\,-\,\omega_0 \,n\,P_n\,+\, 
\omega_0 \,(n-1)\,P_{n-1}. 
\eeq
The first term in the r.h.s. 
can be viewed as a probability of the dipole annihilation 
in the  rapidity range $( y \,\div\, y - dy )$. The second is a
probability to  create one extra  dipole. Note the sign minus
in front of $\partial P_n(y)/\partial y$. It appears due to our choice of the
rapidity evolution which starts at the largest rapidity $y=Y$
of the fastest dipole and then decreases.

It is useful to introduce the generating function \cite{LE,MUUN,LELA} 
\beq   \label{Z}
Z(y,\,u)\,\,=\,\,\sum_n\,\,P_n(y)\,\,u^n\,,
\eeq 
At the rapidity $y\,=\,Y$ 
there is only one fastest dipole, that is $P_1(y\,=\,Y)\,=\,1$
while $P_{n>1}(y\,=\,Y)\,=\,0$. 
This is the initial condition for the generating
function: $$Z(y\,=\,Y)\,=\,u\,.$$ 
At $u =1$ 
\beq \label{INC2}
Z(y,\,u\,=\,1)\,\,=\,\,1
\eeq
which follows from the physical meaning of $P_n$ as  probability.

\eq{PNEQ} can be rewritten as the equation in partial derivatives 
for the generating function $Z(y,u)$: 
\beq \label{GFEQ}
-\,\,\frac{\partial\,Z(y,\,u)}{\partial\, y}\,\,
=\,\,-\,\omega_0 \,u\,(1\,-\,u) 
\,\,\frac{\partial\,Z(y,\,u)}{\partial\, u}
\eeq
The function $Z$ obeying \eq{GFEQ} with the above given initial condition
\beq \label{SOLZ}
Z(Y-y,\,u)\,\,=\,\,\frac{u\,\,e^{\,-\,\omega_0\, (Y - y)}}{1\,\,+\,\,u\,\,
(e^{\,-\,\omega_0 \,(Y\, -\, y)}\,-\,1)}\,.
\eeq

It is important to observe that \eq{GFEQ} can be rewritten in the nonlinear
form. The general form of this equation is such that the solution to this
equation is $Z(u(y))$: when $Z(u(y))$ is substituted into   \eq{GFEQ}, the
derivatives $\partial Z/\partial u$ cancel and we are left with an ordinary
differential equation for $u(y)$. 
Using initial condition at $y\,=\,Y$ we can write
\beq \label{NLEQZ}
-\,\,\frac{\partial \,Z}{\partial \,y}|_{y = Y}\,\,=\,\,
-\,\omega_0 \,\left(\,Z\,\,-\,\,Z^2\,\right)|_{y = Y}\,.
\eeq
Due to  implicit dependence on $y$ this form will be preserved at 
any value of $y$. 
Therefore, the linear equation (see \eq{GFEQ}) can be rewritten 
in the non-linear  form:
\beq \label{GFFD2}
-\,\,\frac{\partial\,Z(y,\,u)}{\partial \,y}\,\,=\,\,-\,\omega_0 
\left(Z  \,\,-\,\, Z^2\right)
\eeq
\eq{GFFD2} coincides exactly (within the same approximation neglecting 
the dipole sizes) with the equation for the generating
functional for the light cone dipole wave function obtained by 
Mueller \cite{MUUN}.
 \eq{GFEQ} can be viewed as a linearized version of the 
non-linear \eq{NLEQZ}.

To find the interaction amplitude   we can follow the procedure suggested in 
Ref.\cite{K}, namely,
\beq \label{AMP}
N_A(y,\,b_t) \,\,=\,\,Im\,A^{el}(y,\,b_t) 
\,\,=\,\,- \sum^{\infty}_{n=1}\,\,\frac{ 1}{n!}\,\,
\frac{\partial^n Z}{(\partial 
u)^n  }|_{u = 1} \,\,\gamma^n(b_t)\,.
\eeq
Here $\gamma(b_t)$ is minus 
imaginary part of the elastic amplitude for a single dipole interaction with  
a nucleus at fixed impact parameter $b_t$. The most important assumption made
in \eq{AMP} is in the independent interaction of $n$ dipoles which is expressed
via $\gamma^n$ factor. In the present work we will relax this
assumption introducing target correlations. 
Note that the amplitude $N_A$ can be found from the following relation
\beq \label{AMP1}
N_A(y,\,b_t) \,\,=\,\,1\,\,-\,\,Z(y,\,\gamma(b_t)\,+\,1)\,.
\eeq

Let us now come back to \eq{GFEQ} from which we are going to derive
a corresponding equation for the amplitude $N_A$.
In order to get  a closed equation, the linear \eq{GFEQ} is differentiated 
$n$ times with respect to $u$ and then set $u=1$:
\beq \label{DLEQGF}
-\,\frac{1}{\omega_0}\,\frac{\partial \,Z^{(n)}}{\partial\, y}
|_{u =1}\,\,=\,\,
n\,\,Z^{(n)}|_{u =1}\,\,+\,2\,\frac{n\, (n - 1)}{2}\,\,Z^{(n-1)}|_{u =1}
\eeq
where $Z^{(n)}\,\,=\,\,\partial^n \,Z/\partial\, u^n$.

The equation for the amplitude can be easily obtained from \eq{DLEQGF}. 
The following four operations are performed. i) Since $Z$ is a function
of  $Y \,-\, y$, we make a substitution 
$\partial Z^n/\partial y\,=\,-\,\partial Z^n/\partial Y$. Hence 
the amplitude evolution will be with respect 
to initial rapidity  of the fastest dipole $Y$. 
(ii) At the rapidity of the target $y\,=\,y_0$ initial conditions for the 
interaction of the 'wee' dipoles with 
the target should be specified.
(iii) For independent interactions with the target (ii) means we
multiply \eq{DLEQGF} by the factor $\gamma^n$. (iv) Divide both sides
by $n!$ and then sum over $n$. 
The result is 
\beq \label{LEQGFA1}
\frac{1}{\omega_0}\,\frac{\partial\,N_A(Y)}{\partial \,Y} \,\,
=\,\, \gamma
\,\frac{\partial\,N_A(Y,\;\gamma)}{\partial\, \gamma} \,\,+\,\,
\gamma^2\,\,\frac{\partial\, N_A(Y,\;\gamma)}{\partial\, \gamma} 
\eeq
\eq{LEQGFA1} 
can be  rewritten in the non-linear form using the 
initial condition $N_A(Y\,=\,y_0)\, =\,-\,\gamma$:
\beq \label{NLEQGFA}
\frac{1}{\omega_0}\,\frac{\partial\,N_A(Y)}{\partial \,Y} 
\,\,=\,\,N_A(Y)\,\,-\,\,N_A^2(Y)
\eeq
\eq{NLEQGFA} is just 
the BK non-linear equation \cite{B,K} in 
our simplified model.

Let us now discuss a way the target correlations can be introduced in 
the analysis. For the linear evolution they appear when we specify in the 
initial  conditions how $n$ 'wee' dipoles interact with the target.
An appropriate modification is to introduce correlation factors $C_n$ 
such that
\beq \label{AMP2}
N_A(y,\,b) \,\,= \,\,- \,\,\sum^{\infty}_{n=1}\,\,\frac{ 1}{n!}\,\,
\frac{\partial^n\, Z}{\partial\, 
u^n  }|_{u = 1}\,\, \,\gamma^n\,\,\,C_n\,.
\eeq 
The coefficients $C_n$ are to be found from a nucleus model. 
The linear equation for the amplitude  now reads
\beq \label{LEQGFA}
\frac{1}{\omega_0}\,\frac{\partial\,N_A(Y)}{\partial \,Y}\,\,
=\,\, \gamma
\,\frac{\partial\, N_A(Y,\;\gamma)}{\partial\, \gamma} \,\,+\,\,
F\left(\gamma\,\frac{\partial}{\partial\, \gamma}\right)
\,\,\gamma^2\,\,\frac{\partial\, N_A(Y,\;\gamma)}{\partial\, \gamma} 
\eeq
Where function $F$ is equal to $F(n) \,=\, C_n/C_{n-1}$ 
and describes  correlations in the 
interactions of $n$ dipoles. In \eq{LEQGFA}, 
$F(\gamma\,\frac{\partial}{\partial\, \gamma})$
should be understood as operator series expansion. In case $F$ is constant
the linear equation (\ref{LEQGFA}) reads
\beq \label{LEQGFAA}
\frac{1}{\omega_0}\,\frac{\partial\,N_A(Y)}{\partial \,Y}\,\,
=\,\, \gamma
\,\frac{\partial\, N_A(Y,\;\gamma)}{\partial\, \gamma}
\,\,+\,\,F\,\,
\gamma^2\,\,\frac{\partial\, N(Y)}{\partial\, \gamma}
\eeq
At $Y\,=\,y_0$ 
the initial condition  is $N(Y\,=\,y_0)\, = \,-\,C_0\,\gamma$. 
It is used to obtain from \eq{LEQGFAA} the non-linear equation: 
\beq \label{NLEQGFAA} 
\frac{1}{\omega_0}\,\,\frac{\partial \,N_A}{\partial \,Y} 
\,\,=\,\,N_A(Y)\,\,-\,\,K_A\, N_A^2(Y)\,.
\eeq
with $K_A\, =\, F/C_0$. In fully non-correlated case $F\,=\,C_0\,=\,1$,
 \eq{NLEQGFAA} reduces to the BK equation for the above
toy model. In the next section we will consider realistic nuclei for
which correlations can be modeled. We will show that within certain 
assumptions the correlation factor $K_A$ is indeed a constant but 
different from unity. It is important to stress that \eq{LEQGFAA} 
can be brought to a nonlinear
form for constant $F$ only. Otherwise $N_A$ is not a function of a single
variable $\gamma(Y)$ and \eq{LEQGFAA} is a linear equation involving
high orders of partial derivatives.

The success of the simple toy model considered above 
leaves us with two hopes for real QCD dynamics.
First, we believe in a possibility to find a linear functional equation  
for the generating functional. A linear equation is more transparent 
and  more feasible to solve compared to the non-linear one.
Second, we believe that \eq{NLEQGFAA} 
(generalized to include BFKL kernel) is the correct generalization 
of the BK non-linear equation applicable 
to targets with correlations such as realistic nuclei.

It turns out that a generalization of a linear functional approach 
to evolution kernels but simplified to a double log accuracy
is rather straightforward \cite{LELA}. 
For the complete BFKL kernel, however,  this is a nontrivial  
problem with which  we will deal below.

\subsection{QCD generating functional for 'fan' diagrams}

In this subsection we consider a linear generating functional
for 'fan' diagrams in QCD. We include both the LO BFKL evolution
of the dipole wave function and QCD triple Pomeron vertex (in large
$N_c$ approximation). Then we essentially  repeat the same derivation 
as for the toy model above. For simplicity of presentation we will 
not treat the impact parameter ($b_t$) dependencies in a correct way
which would imply taking care of all relevant shifts in $b_t$. Instead,
we will consider $b_t$ just as an external parameter which often
will  not be written explicitly. It should be of no problem to reconstruct
a correct $b_t$ dependencies in the final results.

As in Ref. \cite{MUUN} we 
introduce a generating functional $Z$
\beq \label{LD1}
Z\left(Y -y,\,r,\,b_t;\,[u] \right)\,\,\equiv\,\,\sum_{n=1}\,\int\,\,
P_n\left(Y -y,\,r,\,b_t;\,r_1,\,r_2, \dots ,r_i\, \dots ,r_n \right) \,\,
\prod^{n}_{i=1}\,u(\vec r_i) \,d^2\,r_i
\eeq 
where $u(\vec r_i) \equiv u_i $ is an arbitrary function of $r_i$. 
$P_n$ stands for a  probability density to 
find  $n$ dipoles with transverse sizes $r_1,\,r_2,\dots,\,r_i\,\dots\,r_n$ 
and  rapidity $y$ in the wave function  of the
fast moving dipole of the size $r$ and rapidity $Y\,>\,y$. Note that contrary
to Refs. \cite{MUUN,K} we define $P_n$ as a dimensionful quantity. 
The functional (\ref{LD1}) obeys two  conditions:
\begin{itemize}
\item \quad At $y\,=\,Y$, \,\,$P_{n=1}\, =\, 
\delta^2( \vec{r}\,-\,\vec{r}_1)$ 
while $P_{n>1}\,=\,0$. In other words,
\beq \label{LDIN1}
Z\left(Y \,-\,y\,=\,0,\,r,\,b_t;\,[u]\right)\,\,=\,\,u(r)\,\,.
\eeq
\item \quad At $u\,=\,1$ 
\beq \label{LDIN2} 
Z\left(Y \,-\,y,\,r,\,b_t;\,[u=1]\right)\,\,=\,\,1\,.
\eeq
This equation follows from the physical meaning of the functional: sum over
all probabilities equals one.
\end{itemize}
Contrary to the toy model, in QCD the probability to 
survive for a dipole of size $r_i$ is not constant 
but equal to (see Ref. \cite{MUUN} for details)
$$
 \bar{\alpha}_s \,\,\omega(r_i)\,\,=\,\,\frac{ 
\bar{\alpha}_s}{2\,\pi}\,\int_\rho 
\,\frac{r_i^2}{(\vec{r_i}\,-\,\vec{r}')^2\,r'^2}\,d^2 r'\,\,=\,\,
\bar{\alpha}_s \,\,\ln(r_i^2/\rho^2)
$$
with $\rho$ being some infrared cutoff and $\bar\as\,=\,\as\,N_c/\pi$.
The probability for a dipole of the size $r_1\,+\,r_2$ to decay into 
two  with the sizes $r_1$ and $r_2$ is equal to
$$ \frac{\bar{\alpha}_s}{2\,\pi} \,\,\frac{(\vec{r_1} \,\,+\,\, 
\vec{r_2})^2}{r_1^{2}\,r_2^2}
$$
Using these two probabilities we can write down 
the equation for $P_n$:
\begin{eqnarray}\label{LEQPF}
- \,\,\frac{\partial\,P_n\left(Y-y,\,r,\,b_t;\,r_1,\,r_2 \dots r_i 
\dots r_n \right)}{ 
\bar{\alpha}_s\,\partial\, y}\,=\,-\,
\sum^n_{i=1}\,\omega(r_i) \,
P_n\left(Y -y,\,r,\,b_t;\,r_1,\,r_2 \dots r_i \dots r_n 
\right)\nonumber \\
+\,\,\sum^n_{i=1} \,\frac{(\vec{r}_i\,+\, 
\vec{r}_n)^2}{(2\,\pi)\,r^2_i\,r^2_n}\,
P_{n - 1}\left(Y -y,\,r,\,b_t;\,r_1,\,r_2
\dots  (\vec{r}_i \,+\, \vec{r}_n)\dots r_{n-1} \right)
\end{eqnarray}

Let us now introduce two operator vertices
\beq \label{V2} 
V_{1 \rightarrow 1}(r,\,[u])\,\,=\,\,
\bar{\alpha}_s \,\,\omega(r) \,\,u(\vec r)\,\,\frac{\delta}{\delta u(\vec r)}
\eeq
and
\beq \label{V1}
V_{1 \rightarrow 2}(r,\,r',\,[u])\,\,
=\,\,\frac{ \bar{\alpha}_s}{2\,\pi} 
\,\,\frac{r^2}{r'^2\,(\vec{r} - \vec{r}')^2}\,\,
\,u(\vec{r}')\,u(\vec{r} \,-\, \vec{r}')\,\,
\frac{\delta}{\delta u(\vec r)}\,.
\eeq 
The functional derivatives with respect to $u(r)$ play a role 
of an  annihilation operator for a dipole of the size $r$. 
The multiplication by $u(r)$ corresponds to
a creation operator for this dipole.

Multiplying \eq{LEQPF} by the product $\prod^n_{i=1}\,u_i$ 
and integrating over all $r_i$ we obtain the 
following linear equation for the generating functional:
\beq \label{LEQZF}
-\,\,\frac{\partial \,Z}{
\bar{\alpha}_s\,\partial \,y}\,\,=\,\,-\,\,
\int\,d^2 r^\prime\,\,V_{1\rightarrow 1}(r^\prime,\,[u])
\,\, Z\,\,
+\,\,\int\,\,d^2 \,r^{\prime\prime} \,\,d^2\,r' \,\,V_{1\rightarrow 2}
(r^{\prime\prime},\,r',\,[u])
\,\, Z\,.
\eeq
Like in the case of previously considered toy model, $Z$ is a function
of a single variable $u(y)$: if $Z(u(y))$ is substituted into \eq{LEQZF}
we find in the l.h.s. 
$\partial \,Z/\partial \,y\,=\,\int d^2 r\,(\delta Z/\delta\,u(r))\,\,
(\partial\,u(r)/
\partial\,y)$. The integrals $\int d^2 r\,\delta Z/\delta\,u(r)$ 
cancel on both 
sides of the equation and we are left with an ordinary differential 
equation for $u(y)$.
With the help of the initial conditions  
(\eq{LDIN1}), \eq{LEQZF}  can be rewritten in
the non-linear form reproducing the same equation for $Z$ 
as in Ref. \cite{MUUN}:
\beq \label{NLEQZF}
-\,\,\frac{\partial\, Z }{
\bar{\alpha}_s\,\partial \,y}\,\,=\,\,- \,\,\omega(r)\,\,
Z\left(Y\,-\,y,\,r,\,b_t;\,[u] \right)
\eeq
$$
+\,\,\int\,\,
\frac{d^2\,r'}{2\,\pi}\,\,\frac{r^2}{r'^2\,(\vec{r}\,-\,\vec{r}')^2}\,\,
Z\left(Y \,-\,y,\,r',\,b_t;\,[u] \right)\,\,
Z\left(Y\,-\,y,\,(\vec{r}\,-\,\vec{r}'),\,b_t;\,[u] \right)\,.
$$

The amplitude $N_A$ is defined  
similarly to \eq{AMP} (see Ref. \cite{K}):
\beq \label{AMPZ}
N_A(Y,\,r,\,b_t) \,\,=\,\,-\,\, \sum^{\infty}_{n=1}\,\,\frac{ 
1}{n!}\,\,C_n\,\,\int\,\prod^n_{i =1}\,\,\left(d^2 \,r_i \, \,\gamma(r_i;b_t)\,
\, \frac{\delta}{\delta u_i } \right)\,Z\left(Y -y,\,r,\,b_t;\,[u] \right)|_{u = 1}.
\eeq
As above $C_n$ describe  correlations in the $n$-dipole target interaction.
In order to derive an expression for the amplitude $N_A$ we apply 
to both sides of \eq{NLEQZF} $n$ functional derivatives with respect to $u$
and then set $u=1$.
Using the notation $Z^{(n)}\,\,\equiv\,\,\prod^n_{i =1}\frac{\delta}{\delta
u_i } \,Z\left(Y -y,\,r,\,b_t;\,[u] \right)$ we obtain
$$
-\,\frac{\partial}{ 
\bar{\alpha}_s\,\partial\,y}\,Z^{(n)}|_{u=1}\,\,=\,\,\left(
-\,\sum_{i=1}^n
 \,\,\omega(r_i)\,\,Z^{(n)}|_{u=1}\,\,\,+\,\,2\,\sum_{i=1}^n\,
\int\,\frac{d^2\,r'}{2\,\pi}\,\,
\frac{r'^2}{r^2_i\,(\vec{r}_i\,-\,\vec{r}')^2}\,\,Z^{(n)}|_{u=1}\,\, \right.
$$
\beq \label{DFZ} 
\left.
+\,\,\sum_{i,\,j=1;\,j\neq i}^n\,\frac{(\vec{r}_i + \vec{r}_j)^2}
{(2\,\pi)\,r^2_i\,r^2_j}\,\,Z^{(n)}|_{u=1}\,\right)\,.
\eeq
The evolution  equation for the amplitude is obtained through the same 
steps as in the toy model above.  We divide \eq{DFZ} by $n!$  and 
multiply by the product $-\,C_n \,\prod^n_{i=1}\,\gamma(r_i,\,b_t)$.  
Then we integrate over all $r_i$ and  sum over $n$. Finally we consider
the evolution with respect to $Y$ instead of $y$ which implies a change
of sign in front of the derivative:
\beq \label{AMPLEQZ}
\frac{\partial\, N_A(Y;\,[\gamma])}{\bar \alpha_s\,\partial\,\,Y}  
\,\,=\,\,- \int\,d^2\,r' \,\,V_{1\rightarrow 1}(r',\,[\gamma(r')])\,\,
N_A(Y;\,[\gamma])\,\,
\eeq
$$
+\,\,\frac{2}{2\,\pi}\,\int \, d^2 \,r'\,d^2 \,r" \,
\,\gamma(r")\,\frac{r'^2}{\,r"^2\,(\vec r"\,-\,\vec r')^2}\,\,
 \frac{\delta}{\delta\, \gamma(r')}
\,\,N_A(Y;\,[\gamma ])
$$
$$
+\,\,F\left(\int \,d^2\,\bar r 
\,\gamma(\bar r)\,\frac{\delta}{\delta  
\gamma(\bar r)}\right)\,\,
\int \,d^2\,r"\,\,d^2\,r'
 \,\,V_{1\rightarrow 2}(r',\,r",\,[\gamma(r')])\,\, 
N_A(Y;\,[\gamma ])
$$
\eq{AMPLEQZ} is linear but in functional derivatives and it is not easy 
to develop  direct methods to solve it. 
In  case of function $F$ being a constant, 
\eq{AMPLEQZ} can be easily reduced to a non-linear equation.  
Using the initial condition $N_A(Y\,=\,y_0)\,=\,-\,C_0\,\gamma$ we
obtain the following nonlinear equation:
\beq    \label{NLEQRA}
\frac{\partial\,N_A(r,\,y;\,b_t)}{\partial\,Y}\,=\,
\bar\as\,\times
\int_{\rho} \, \frac{d^2\,r'}{2\,\pi}\,
\frac{r^2}{r'^2\,(\vec r\,-\,\vec r')^2}\,\,\times
\eeq
$$
\left[\,2\,N_A(r',\,y;\,b_t)
 \,\,-\,\,N_A(r,\,y;\, b_t)\,\,-\,\,K_A\,\,N_A(r',\,y;\,b_t)\,\,
N_A(\vec r\,-\,\vec r',\,y;\,b_t)  
\right]\,, $$
with $K_A\,\equiv\,F/C_0$. 
If we consider dipole interaction as fully uncorrelated $F\,=\,C_0\,=\,1$
then \eq{NLEQRA} reduces to the BK equation. We will find below
that within some assumptions, correlations in realistic nuclei can be described
by a constant $F$, but with $F\,\ne \,C_0$. In this case \eq{NLEQRA} is
a generalization of the BK equation to realistic nuclei with 
correlations.

\section{Dipole-nucleus interactions with correlations}

\subsection{Nucleus correlations}
Let us  discuss the target correlations starting with dipole-nucleus
interactions at  sufficiently low energy ($y_0 =\,\ln(1/x_0)$). 
It should be stressed that 
the energy is supposed to be not too 
low such that high energy QCD is applicable. On the other hand, we assume
that energy is low enough such that a single dipole interacts with
one nucleon only. We neglect all Glauber rescatterings at this energy.

\begin{figure}
\begin{center}
\epsfig{file=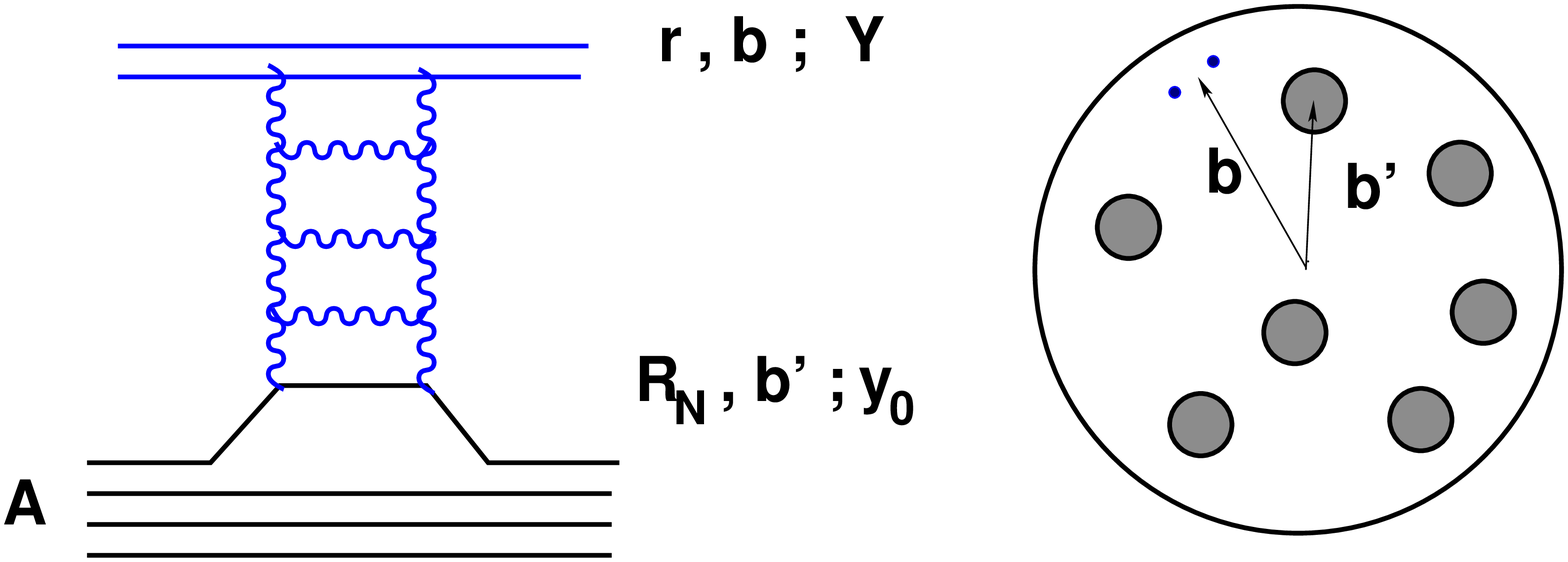,width=150mm}
\end{center}
\caption{The exchange of the single  BFKL Pomeron 
for dipole-nucleus scattering.}
\label{leq}
\end{figure}

\begin{figure}
\begin{center}
\epsfig{file=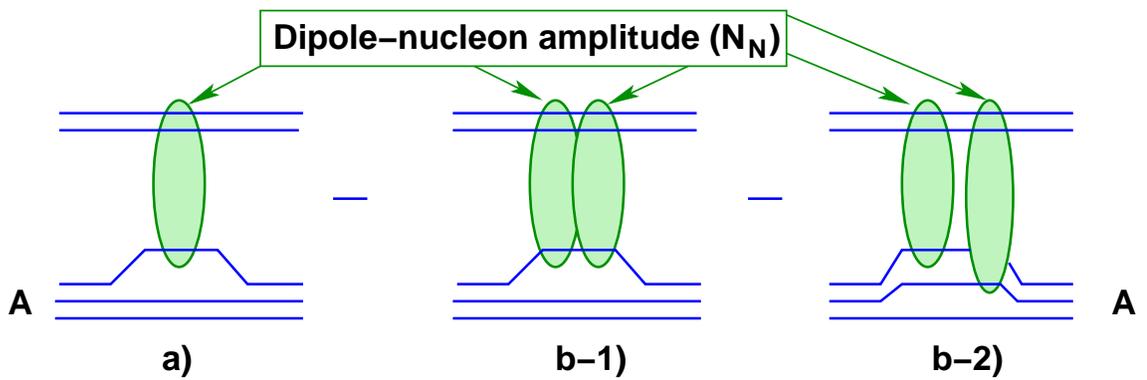,width=150mm}
\end{center}
\caption{A single dipole and two dipole nucleus amplitudes.}
\label{ba}
\end{figure}

Consider first the \fig{leq} which presents in details how a single dipole
interacts with a nucleus. Schematically the same is 
displayed in (\fig{ba}-a).  
Its  contribution is proportional to $\kappa_A(b_t)$: 
\beq \label{NANN}
\int d^2  b'\,
N_N(r,\,y_0;\,\vec{b}_t \,-\, \vec{b}_t') 
\,S_A(b_t')\,= 
\,S_A(b_t)\,\int d^2 b_t"\,N_N(r,\,y_0;\,b_t")\,\approx\,
\,N_N(r,\,y_0;\,b_t=0)\,\kappa_A(b_t)\,.
\eeq
Here $N_N$ stands for the imaginary part of the dipole-nucleon elastic
amplitude. $N_N$ can, in principal, include multiple dipole-nucleon
rescatterings. As it is clear from \eq{NANN} and \fig{leq} that $b_t$
is a distance from the nucleus center to a given nucleon.
  
In \eq{NANN} the first equality is due to the 
assumption, standard for the Glauber theory, that the size of the 
nucleon as  well as the radius of the dipole-nucleon interaction is much 
smaller than the size of the nucleus. The second equality is model dependent. 
We have assumed a cylindrical nucleon that is 
$N_N(r,\,y_0;\,b_t)\,=\,N_N(r,\,y_0;\,b_t =0)\,\,\Theta(R_N \,-\, b_t)$. 
The parameter $\kappa_A(b_t)$ (\eq{KAPPA}) measures the strength of 
the dipole nucleus interaction.

Consider now diagrams of \fig{ba}-b. 
For very large nuclei the diagram of \fig{ba}-b-2 dominates since it is 
proportional to $\kappa^2_A(b_t)$ while the  diagram of \fig{ba}-b-1 
is of the order  $\kappa_A(b_t)$ only. 
In the case when $\kappa_A(b_t)$ is not  large  the  diagram of \fig{ba}-b-1
should be also taken into account. 
Its contribution  has the form:
\beq \label{BA1}
-\,
\int\,\,d^2\,b_t'\, \left(N_N(r,y_0;\vec{b}_t - \vec{b}_t')\right)^2 
\,S_A(b_t')\,\,
\simeq\,\,\left(N_N(r,\,y_0;\,b_t\,=\,0)\right)^2 \,\,\kappa_A(b_t)\,.
\eeq
\fig{ba}-b should be viewed as interaction of two dipoles with the nucleus.
\fig{ba}-b-1 is a correlated interaction with one and the same nucleon while
\fig{ba}-b-2 is independent interaction with different nucleons. 
In \eq{BA1} we have neglected the difference in the dipole sizes, which 
is not important for this discussion.

If we proceed further for higher order terms
we can find a general expression for $n$-th term in the expansion.
For three dipoles we have a term proportional to $\kappa_A$ when all
three interact with the same nucleon, a term proportional to  $\kappa_A^2$
when two dipoles are correlated, and a term proportional to  $\kappa_A^3$
when all interact independently. The sum of these terms will be
$$
N_N^3\,\,
(\kappa_A\,+\,2\,\kappa_A^2\,+\,\kappa_A^3)\,.
$$
The factor 2 in front of $\kappa_A^2$ is due to combinatorial counting.
The $n$-th term in the expansion has the form:
\beq \label{BA2}
(-1)^n \,\,\kappa_A\,(1 \,+\, \kappa_A)^{n - 1}\,\,N_N^n\,.
\eeq
There is no time ordering associated with the factor $1/n!$ which usually
appears in the Galuber expansion.
$(-1)^n$ stems from the fact that the dipole-nucleon amplitude at 
$y_0 \,=\, \ln(1/x_0) \,\gg\,1$ is pure imaginary. 

From \eq{BA2} we can read off the correlation coefficients $C_n$ introduced
above:
$C_n\,=\,\kappa_A\, ( 1 \,+\, \kappa_A)^{n -1}$. Hence the correlation function
$F\,=\,( 1 \,+\, \kappa_A)$  and 
the correlation parameter $K_A\,=\,(1\,+\,\kappa_A)/\kappa_A$. The correlations
appeared when the assumption about  independent interactions was relaxed. 
In spite of the introduction of the correlation parameter,
\eq{BA2} still allow interpretation in terms
of independent dipole-nucleus interactions. The strength of this interaction is
$(1\,+\,\kappa_A)\,N_N$. In fact the dipoles interact 'almost' 
independently and only the overall factor  $\kappa_A/(1+\kappa_A)$ incorporates
the correlations inside the nucleus. The most
important fact is that we were able to introduce only one correlation 
parameter independent of the number of dipoles participating
in the interaction. This success can be traced back to the model assumption
about the cylindricity of nucleons. 

\subsection{Nonlinear equation for realistic nuclei}

Above we have considered correlated interactions
of $n$-dipoles  with a realistic (relatively dilute) nucleus.
We have found  the correlation factor $K_A\,=\,(1\,+\,\kappa_A)/\kappa_A$
which is to be substituted into \eq{NLEQRA}. As a result,  \eq{NLEQRA}
gives a generalization of the BK equation to realistic nuclei
with correlations.  As was already mentioned in 
the introduction, \eq{NLEQGFAA} respects both small and large limits of 
$\kappa_A$. It is most important to stress that the correlation factor $K_A$
is always $\gg \,1$. This means the nucleus correlations increase the
shadowing. In other words, the BK equation  which assumes independent
interactions underestimates shadowing corrections for realistic nuclei.

Let us now present some numerical estimates related to  \eq{NLEQGFAA}.
The effective nucleon radius found in Ref. \cite{LUB} is
$R_N^2\,=\,3.1\,GeV^2$. At zero impact parameter $b_t\,=\,0$, $\kappa_A(b_t)$
is maximal. For the heaviest nuclei, say gold ($A=197$), 
$\kappa_{Au}(0)\,\simeq\,0.85$  while for neon ($A=20$) 
$\kappa_{Ne}(0)\,\simeq\,0.35$.

\begin{figure}
\center\epsfig{file=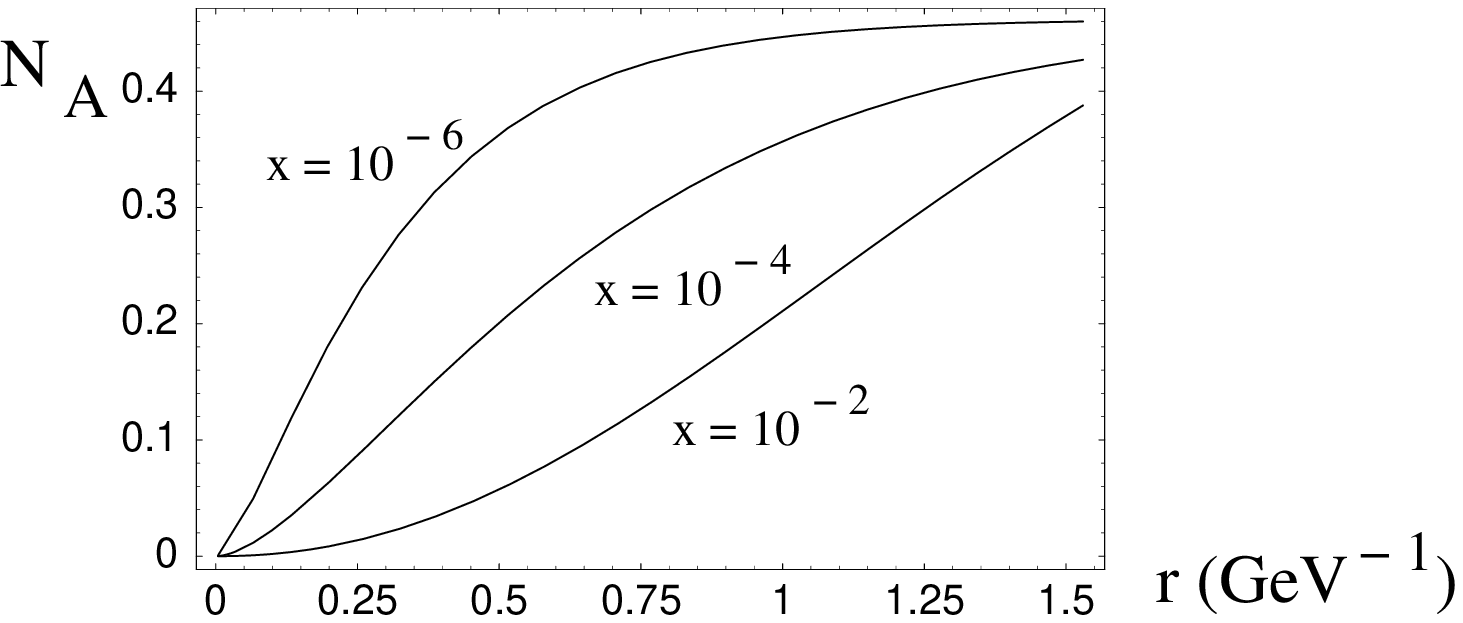,width=100mm}
\caption{The solution of \eq{NLEQGFAA} as a function of $r$ for $b_t\,=\,0$.}
\label{solBK}
\end{figure}

In the deep saturation region a solution of \eq{NLEQGFAA} will saturate.
The saturation bound will not be unity but $N_A \,=\,1/K_A\,\le\,1$. 
In order to confirm our expectations 
\eq{NLEQGFAA} is solved numerically for gold nucleus
at $b_t\,=\,0$ (\fig{solBK}). 
The initial conditions are set at $x\,=\,x_0\,=10^{-2}$.
Consistently with our derivation we take 
$N_A(x_0,\,r)\,=\,\kappa_A(0)\,N_N(x_0,\,r)$ with $N_N$ being a 
dipole nucleon amplitude found in Ref. \cite{LUB}. The function $N_A$
shown in \fig{solBK} indeed saturates  at the value $1/K_A$ in accord with
the expectations. 

\subsection{Non-linear equation and enhanced diagrams}.

\begin{figure}
\begin{tabular}{c c}
\epsfig{file= 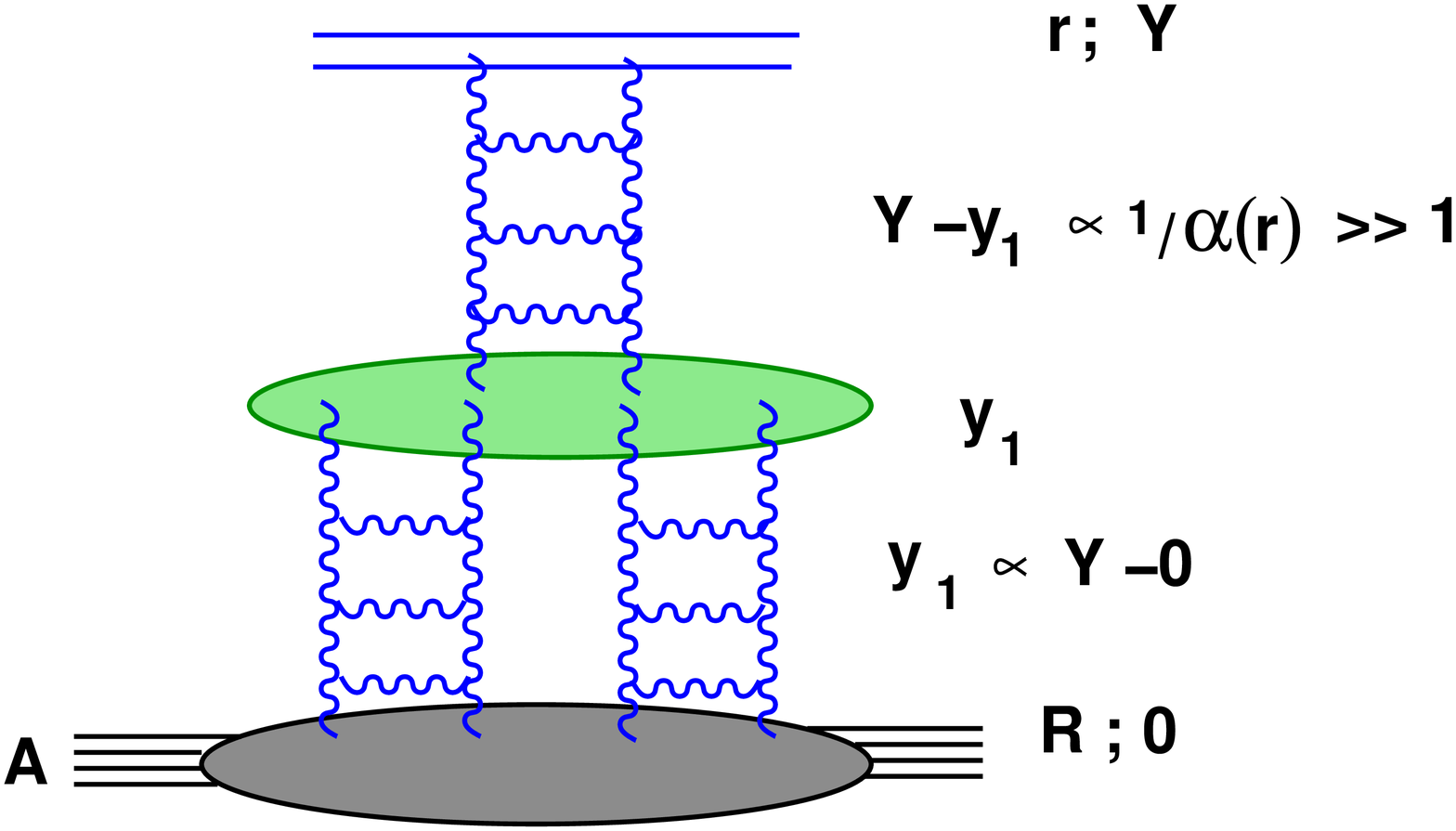,width=75mm}&
\epsfig{file= 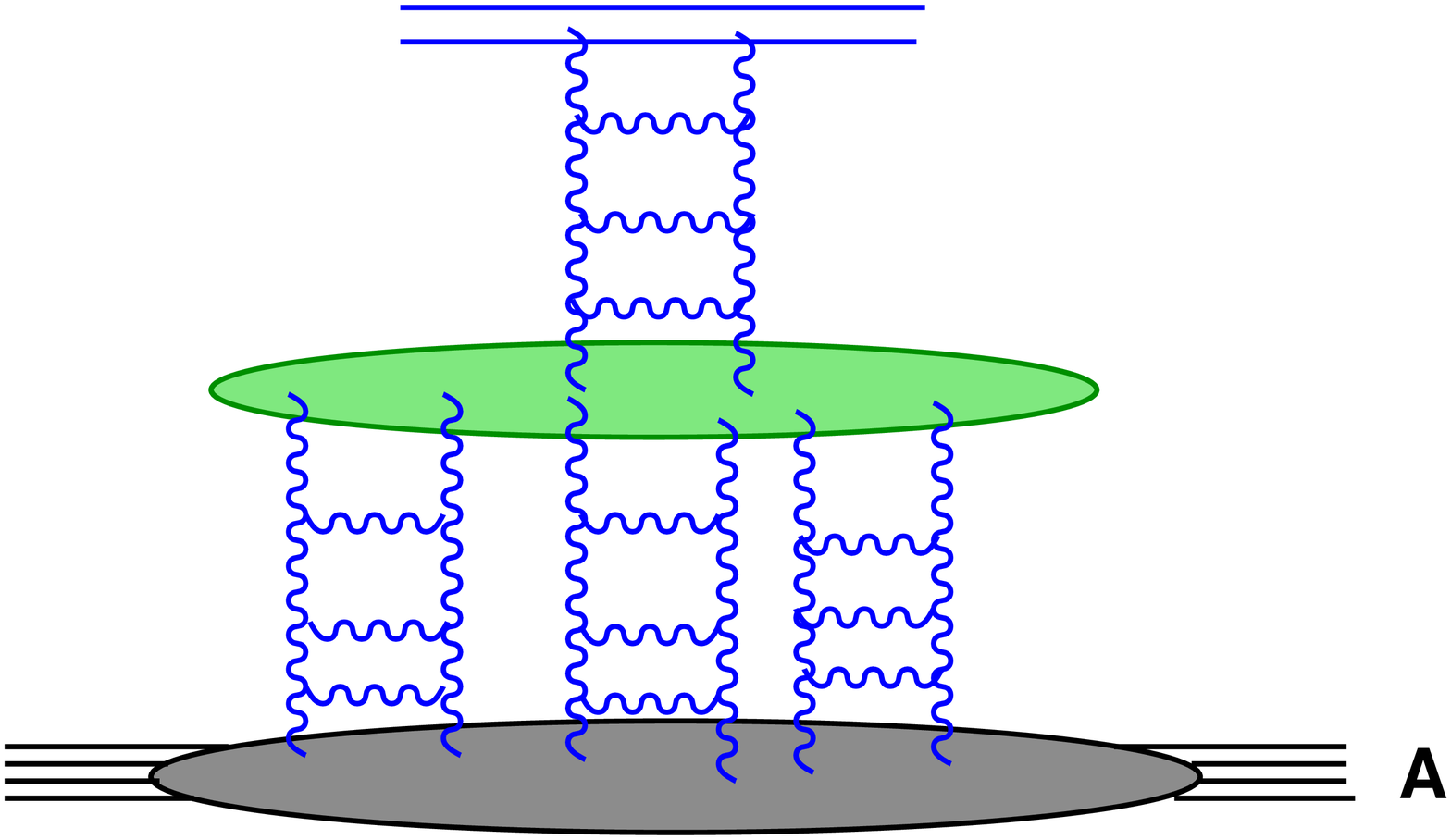,width=65mm} \\
       &    \\
\fig{endi}-1 & \fig{endi}-2 \\
   &    \\ 
\epsfig{file= 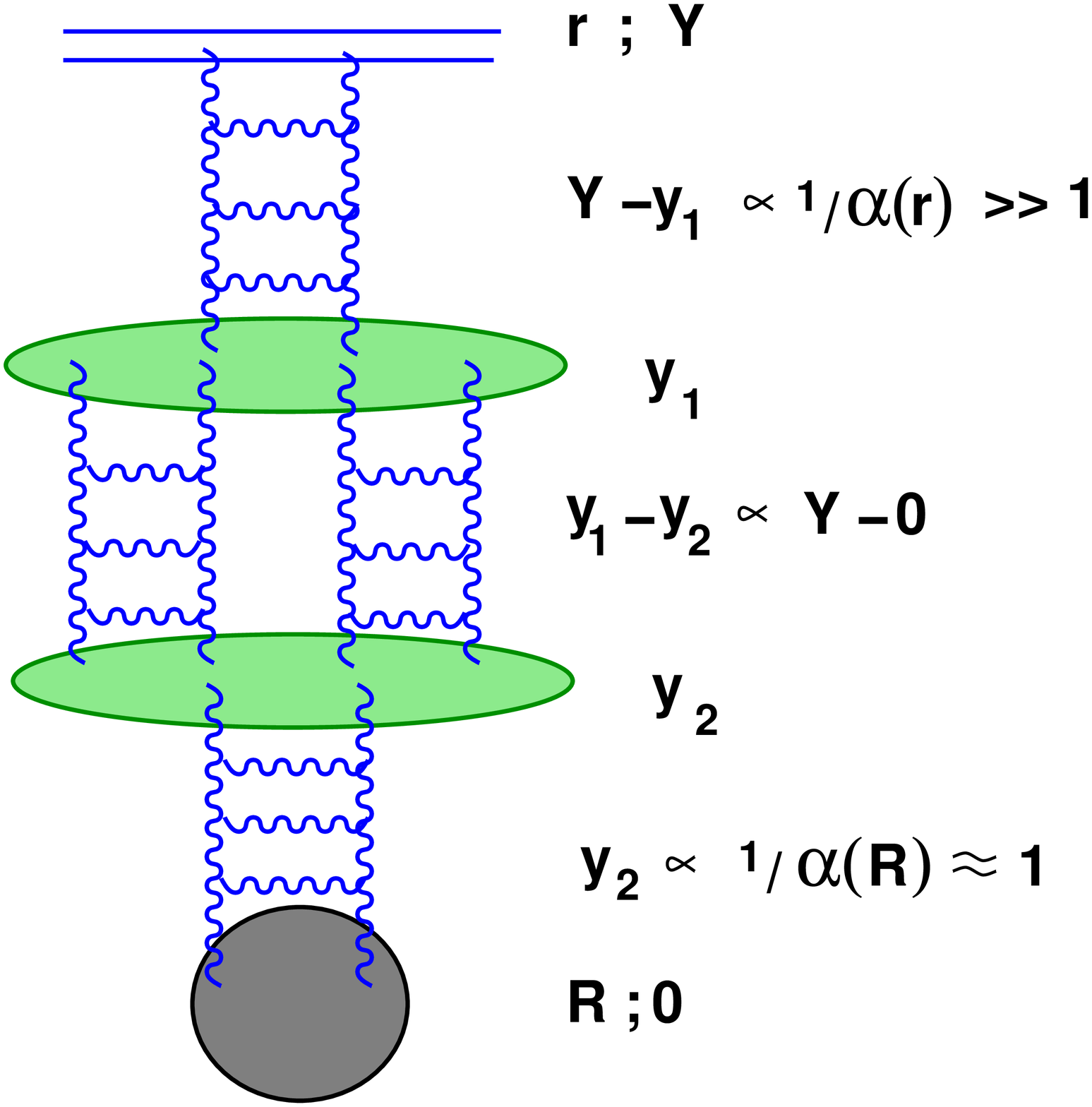,width=60mm}&
\epsfig{file=  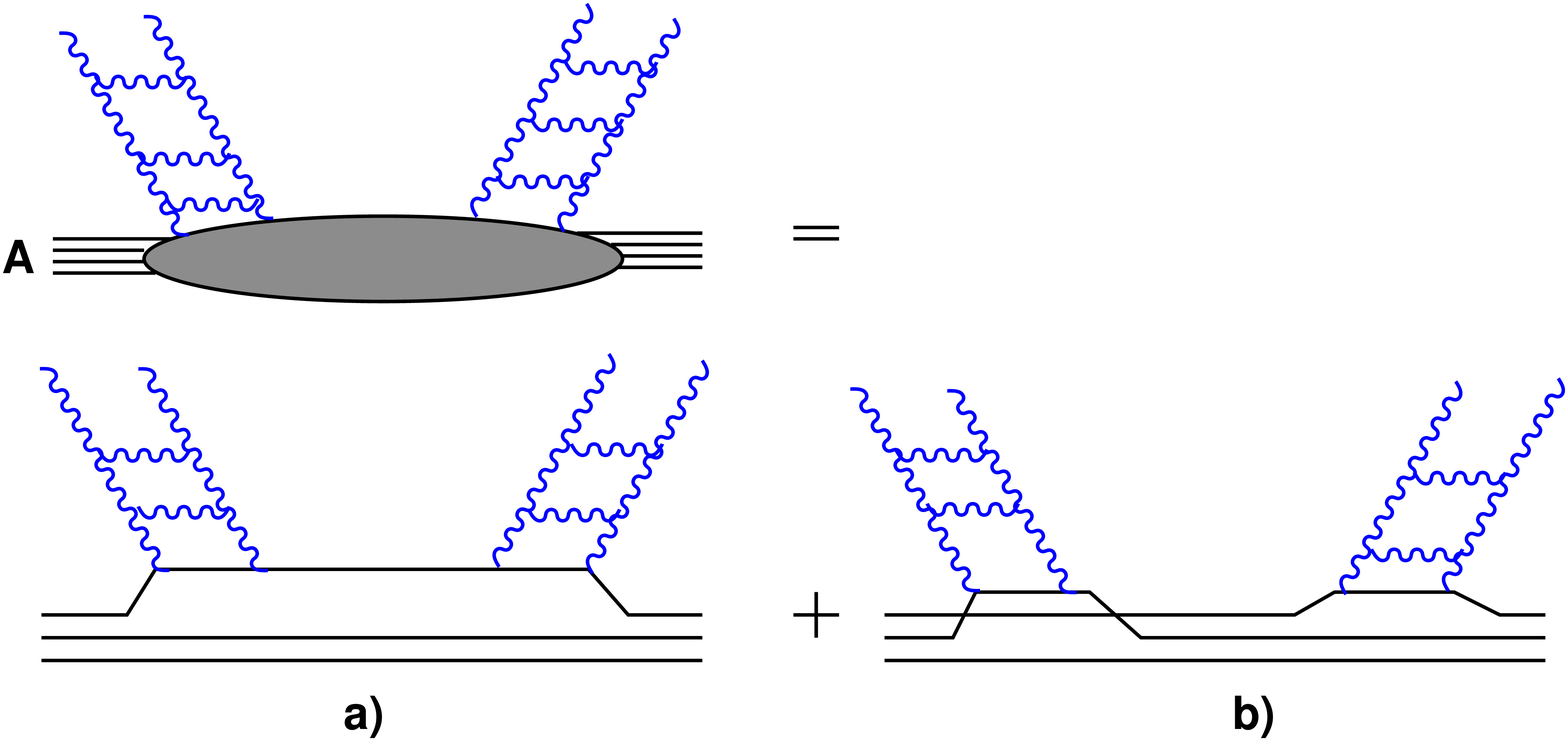,width=80mm} \\
       &    \\
\fig{endi}-3 & \fig{endi}-4 \\
   &    \\
\end{tabular}

\caption{The 'fan' diagrams (\fig{endi}-1 and  \fig{endi}-2)
for dipole-nucleus interaction.
 First enhanced diagram (\fig{endi}-3).
\fig{endi}-4 shows a model for the 
interaction of the BFKL Pomeron with  nucleus target.}
\label{endi}
\end{figure}

The BK nonlinear evolution equation  sums the 'fan' diagrams of \fig{endi}-1.
It is important to understand the limits of its applicability and develop 
methods to include necessary corrections. Two types of diagrams which are not
included in the BK equation are show in \fig{endi}-2 and \fig{endi}-3. 
These are semi-enhanced diagrams including multiple Pomeron vertices 
(\fig{endi}-2) and the enhanced diagram with Pomeron loops. 

The diagram of \fig{endi}-2 can be neglected in pre-saturation region only,
when $N_A \ll 1$. Thus, in spite of its importance to high energy 
phenomenology, the validity of the BK equation is limited to 
$r\,<\,1/Q_s$. In the next section we will propose a new non-linear equation,
which is supposed to sum the 'fan' diagrams  as well as those of 
\fig{endi}-2. 

The assumption of large $\kappa_A \,\gg\,1$ was essential in arguments given
in Ref. \cite{K} in order to neglect the enhanced diagrams of \fig{endi}-3.
Indeed, the 'fan' diagrams dominate in this limit \cite{SCHW}. 
As we have argued above,
for realistic nuclei $\kappa_A \,\le\,1$ and the argument is not applicable.
Nevertheless we are still convinced that the Pomeron loop diagrams are 
subleading provided the projectile dipole has a small size.

 Following  Refs. \cite{GLR,MUQI,MV} we can estimate a contribution of the
enhanced diagram of \fig{endi}-3. 
The ladders are associated with the BFKL Pomeron having
the property discussed earlier:
\beq \label{NLE1}
N \,\,\propto\,\,e^{\omega_0 \,y}
\eeq
with $\omega_0\,\propto\,\alpha_s$.
The integration over $y_1$ and $y_2$  leads to
\beq \label{NLE2}
N^{enhanced}_A\,\,\,\,\propto\,\,\,\,\frac{1}{\omega_0(r)\,\,\omega_0(R)}\,\,
\,e^{\,2\,\, \omega_0\, (Y \,-\,0)}\,\,\approx\,\,
\frac{1}{\alpha_s(r)\,\alpha_s(R)} \,\,e^{2\,\,\omega_0\, Y}\,.
\eeq
The typical values of $Y - y_1 \,\approx\,1/\omega(r)\,\,\gg\,1$ while 
$y_2 - 0 \,\,\approx\,\,1/\omega(R)\,\,\,\approx\,\,1$.
Therefore, the upper ladder which is  in the region 
where the QCD coupling is small, describes a 
high energy process for which the BFKL equation can be trusted.
On the other hand, the lowest ladder corresponds to the interaction at 
rather low energies where high energy asymptotic expressions are not
applicable. Consequently, the enhanced diagram of \fig{endi}-3 should be
viewed as degenerate with the 'fan' diagram of \fig{endi}-1.
A more accurate estimate of \fig{endi}-3 involves  integrations over 
dipole sizes  in the vertices but they do not alter the conclusion (see
Refs. \cite{GLR,MUQI} for details). The argument above depends crucially
on the running QCD coupling and the fact that $\alpha_s(r)\,\ll\,\alpha_s(R)$.
Concluding we expect the equation  summing  'fan' diagrams only
be justified for nucleon targets if the projectile dipole  $r$ is small,
such that $\alpha_s(r) \,\ll\,1$.

It is worth to emphasize that even the `fan' diagram contribution depends on 
the model for the `two ladders-nucleus' vertex (see \fig{endi}-1). 
In this paper we have assumed a model which  graphically is represented 
in  \fig{endi}-4. This model includes the eikonal-type of 
interaction with the nucleon.

\section{Glauber rescattering and QCD generating functional}

As was discussed above,  \eq{LEQZF} for generating functional, which is 
equivalent to the one obtained by Mueller, describes the 'fan' diagrams only.
The fact that this equation does not describe the semi-enhanced diagrams of 
the type shown in \fig{endi}-2 is an apparent shortcoming. At dipole sizes
$r \,\geq\,R_s(x)\,=\,2/Q_s(x)$ these diagrams are of the same order as the
'fan' diagrams and they have to be taken into account in order to  obtain 
a reliable description of the saturation domain. In this section we propose
a generalization of the QCD functional approach to such type of interactions.
To simplify the presentation  we will not consider  target correlations 
in this section.

In order to solve the problem we have to find an effective
 vertex for one dipole splitting
to many. This vertex can be deduced from the Glauber-Mueller formula 
\cite{MU90,AGL,MUKO} which
describes  propagation of a dipole through large nucleus. It is important
to stress that this formula was proven in QCD. In our approach,
the effective vertex will be obtained as a chain of multiple rescatterings
involving a dipole splitting to two dipoles.

The Glauber-Mueller formula reads
\beq \label{GLUPRO}
N^{GM}(y,\,r,\,b_t)\,\,=\,\,
\left(\,1\,\,-\,\,e^{- N^{BFKL}(y,\,r,\,b_t)}\,\right)
\eeq
with $N^{BFKL}$ being a solution to the linear BFKL equation. We can use
\eq{GLUPRO} to define initial conditions for a new functional.
 $N^{BFKL}$ describes  a single Pomeron exchange which in the language of
section 2 should be replace by $1\,-\,u$. Basing on the relation \eq{AMP1}  
we propose the following initial conditions at $y \,=\, Y$: 
\beq \label{INCEK1}
Z\left(Y-y\,=\,0,\,r,\,b_t;\,[u]\right)\,\,=\,\,e^{u(r)\,-\,1}\,\,.
\eeq
At $u\,=\,1\,\,$,  $\,Z$ has the correct normalization: $Z\,=\,1$.

\eq{INCEK1} can be given the following interpretation. Contrary to the case
considered in section 2 where we had a single dipole at the beginning
of the evolution, \eq{INCEK1}  describes  initial state in which many dipoles
can be created. $Z$ represents a probability distribution for a dipole number
in the initial state. It can be rewritten as
\beq \label{INCEK1A}
Z\left(Y\,=\,y,\,r,\,b_t;\,[u]\right)\,=\,
\sum_{n =0} e^{\, -\, <n>}\,\frac{(u\,<n>)^n}{n!}
\eeq
The average number of dipoles 
$<n>\, \equiv\,\frac{\delta}{\delta \,u} Z|_{u=1}\,=\,1$.
\eq{INCEK1A} is a typical Poisson distribution with 
$e^{\, -\, <n>}\,<n>^n/n!\,=\,e^{-1}/n!$ being a probability to find
$n$ dipoles (of the same size) in the initial state.
 
As in section 2, the overall probability conservation implies that
  at $u \,=\,1$
\beq \label{INCEK2}
Z\left(Y-y,\,r,\,b_t;\,[u=1]\right)\,\,=\,\,1\,
\eeq
to be fulfilled at any rapidity.

We are going now to 
find  a new effective vertex $\mathcal{V}$
for one dipole splitting to many. The strategy
is following. We consider a two step process. First, a dipole of size $\vec r$
decays into two dipoles of sizes $\vec r'$ and $\vec r\,-\,\vec r'$. This
process is described by the vertex $V_{1 \rightarrow 2}$ introduced in \eq{V1}.
Then each  produced dipole is considered not as a single dipole but
rather having a Poisson distributed dipole multiplicity of equal size dipoles.
This step is described using the Glauber - Mueller formula.  

The amplitude $\tilde N$ for this two step propagation 
can be written in the form \cite{MUKO}
\beq \label{GM}
\frac{\partial\,\tilde{N}(y,\,r)}{\partial y}\,\,
=\,\,- \,\,\bar{\alpha}_s\,\,\omega(r)\,\, \left( 1 
\,-\, e^{ \,-\, N^{BFKL}(y,\,r)} \right)\,\,+
\eeq
$$
\,\,\frac{\bar{\alpha}_s}{2\,\pi}\,\,\int\,\frac{r^2\,  
d^2 r'}{r'^2 \,(\vec{r}\,-\,\vec{r}')^2}\,\left( 1\,\,-\,\,
e^{\, - \,\{ N^{BFKL}(y,\,r')\,\,+\,\,N^{BFKL}(y,\vec{r}\,-\,\vec{r}')\,\} 
}\right)\,.
$$
The first term in \eq{GM} is a 
possibility for the parent dipole ($r$) not to decay while the second term
describes the decay. The  probability for this decay is given by the square
of the wave function 
$|\Psi|^2\,=\,\bar{\alpha}_s\,\frac{r^2 }{r'^2\,(\vec{r}- \vec{r}')^2}$. 
Two fast dipoles thus created then percolate through 
the target preserving their sizes. This process is described by \eq{GM}. 
 
It is important to explain the difference between \eq{GLUPRO} and \eq{GM}.
\eq{GLUPRO} describes a propagation of  a dipole with the size $r$ 
through target without decay. Alternative it can be viewed as including
decays but with decay products being identical dipoles with the same size
and rapidity. On the other hand, \eq{GM} describes the two stage process. 
The physics of \eq{GM}  could be clarified if we rewrite it in the following
form
\beq \label{GMN}
\frac{\partial\, \tilde{N}(y,\,r)}{\partial y}\,\,=\,\,- \bar{\alpha}_s 
\,\omega(r)\,N^{GM}(y,\,r)\,\,+\,\,
\eeq
$$
\frac{\bar{\alpha}_s}{2\,\pi}\,\,\int\,\frac{r^2\,d^2 r'}{r'^2\, 
(\vec{r}\,-\,\vec{r}')^2}\,\,\left(\,N^{GM}(y,\,r')\,\,+\,\,
N^{GM}(y,\,\vec{r}\,-\,\vec{r}')\,\,- 
\,\,N^{GM}(y,\,r')\,\,N^{GM}(y,\,\vec{r}\,-\,\vec{r}') 
\right)\,.
$$
We would like to emphasize that the splitting kernel is the same LO BFKL
kernel as appears in the  BK equation.

We are now in the position to extract from \eq{GM} an effective vertex 
$\mathcal{V}$. We interpret the Glauber- Mueller formula as 
an annihilation of a scattering dipole and a creation of a 
number of identical new dipoles. Basing on our previous experience in section 2
we substitute in the r.h.s. of \eq{GM}  $1 \,-\, u$ instead of $N^{BFKL}$ 
(see \eq{AMP} and Ref.\cite{K}) and  obtain the generalized vertex
\beq \label{VG}
\mathcal{V}(r;\,[u])\,\,
=\,\,\bar{\alpha}_s\,\left( -\,\, \omega(r)\,\,e^{u(r) - 
1}\,\,+\,\,\int\,\frac{r^2 \, d^2 r'}{2\,\pi\, r'^2\,(\vec{r}- \vec{r}')^2} 
\,\,e^{ \,u(r')\,-\, 1\,\, +\,\, 
u( \vec{r}- \vec{r}') \,-\, 1}\,\right)\,\,\frac{\delta}{\delta\, u(r)}\,\,
\eeq
The vertex $\mathcal{V}$  being expanded in $u\,-\,1$ reproduces
the vertices $V_{1\rightarrow 1}$ and  $V_{1\rightarrow 2}$ as a correct
limit.

Having obtained the vertex \eq{VG}, the linear equation for the generating 
functional (see \eq{LEQZF}) can be generalized straightforwardly:
\beq \label{LEQZK} 
-\,\,\frac{\partial\, Z}{\partial\,y}\,\,\,\,=\,\,\,\,
\int \,\,d^2\,r'\,\,\,\mathcal{V}(r';\,[u])\,\,\,Z\,.
\eeq
As was mentioned above, if the vertex  $\mathcal{V}$ is expanded in $u\,-\,1$,
 \eq{LEQZK} reduces to \eq{LEQZF} and sums the diagrams of \fig{pneq}.

Following the same steps as in section 2, \eq{LEQZK} can be reduced to
a non-linear equation. At $Y\,=\,y$ the initial condition of \eq{INCEK1} 
lead to  the relation
\beq \label{RHSEQ}
\frac{\delta\,Z}{\delta\,u}\,\,\,=\,\,\,Z\,.
 \eeq
Since \eq{LEQZK} can in fact be rewritten as ordinary differential
equation, we can use the relation (\ref{RHSEQ}) at any rapidity.
This results in the new non-linear equation for the functional
\begin{eqnarray} \label{NLEQZK}
&-&\frac{\partial\,\ln\, Z\left(Y -y,\, \vec{r},\,b_t;\,[u]\right)}{
\bar{\alpha}_s\,\partial\,y}\,\,=
 \,\,-\,\,\omega(r) \,\,Z\left(Y -y,\, \vec{r},\,b_t;\,[u]\right)\,\\
& &+\,\,\int\,
\frac{d^2\,r'}{2\,\pi}\,\,\frac{r^2}{r'^2\,(\vec{r}\,-\,\vec{r}')^2} \,\, 
Z\left(Y -y, \vec{r}',b_t;\,[u]\right)\,\,
\times\,\,  Z\left(Y -y,\, \vec{r}\,-\,\vec{r}',\,b_t;\,[u]  \right)\,.
\nonumber
\end{eqnarray}
\eq{NLEQZK} is a new equation summing up not only the 'fan' diagrams
 of \fig{endi}-1 but also  of the type displayed in \fig{endi}-2. \eq{NLEQZK}
compared to \eq{NLEQZF} has an extra power of $Z$ which we 
are not able to explain using a physical intuition. Nevertheless we believe
\eq{NLEQZK} is correct in describing all types of 'fan' diagrams including
multiple Pomeron vertices. This equation has a very good chance to 
describe the saturation region. As we have argued that inside of 
the saturation region all diagrams of semi-enhanced type are essential.

Let us finally write down a new non-linear equation 
for the scattering amplitude defined in  
\eq{AMPZ}. Using the relation \eq{AMP1} between the generating functional
$Z$ and the amplitude, the equation for the latter reads
\beq \label{NLEQNK}
\frac{ \partial\,N_A\left(Y,\,r;\,b_t\right)}{
\bar{\alpha}_s\,\partial\,Y}\,\,=
\,\, \left[ 1\,-\, N_A(Y ,\,r;\,b_t) \right] \times
\eeq
$$
\int\,\,\frac{d^2\,r'}{2\,\pi}\,\,\frac{r^2}{r'^2\,(\vec{r}\,-\,\vec{r}')^2} \,
[\,
2\,\,  N_A(Y,\,r';\,b_t)\,\,-\,\,N_A(Y,\,r;\,b_t)
-\,\, 
N_A(Y,\,r';\,b_t) \,\times \, N_A(Y,\,
\vec{r}\,-\,\vec{r}';\,b_t)\,]  \,.
$$
The initial conditions for \eq{NLEQNK} are the Glauber-Mueller formula
$N_A(Y\,=\,y_0)\,=\,N^{GM}(y_0)$. \eq{NLEQNK} has the correct 
linear limit (LO BFKL) when $N_A$ is small. There is a region, which is not
defined by any expansion parameter, where $N_A\,\ll\,1$ but $N_A^2$ should not
be neglected compared to $N_A$. In this region \eq{NLEQNK} reduces to the 
BK equation.

\subsection*{Quantitative estimates}

\begin{figure}[htbp]
\begin{minipage}{10.0cm}
\epsfig{file= 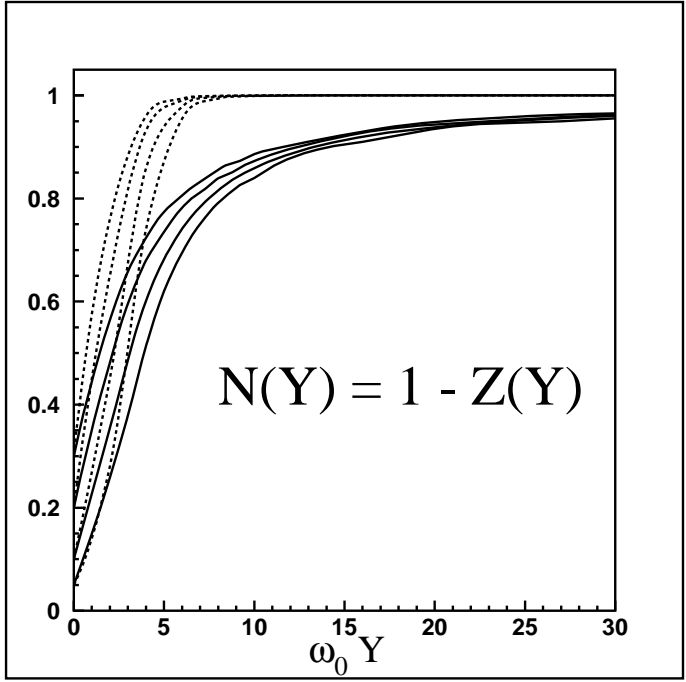,width=90mm}
\end{minipage}
\begin{minipage}{6.0 cm}
\caption{The amplitude obtained in the toy model.
The dotted lines show the solutions of the BK equation while the solid curves 
correspond to the solution of \eq{TMZSOL}. Four different values for initial
conditions are plotted.}
\label{sol}
\end{minipage}
\end{figure}

In order to estimate the impact of the newly accounted diagrams of
\fig{endi}-2 we again consider the simple toy model of section 2.1.
In this model  \eq{NLEQZK} reads
\beq \label{TMNEQ}
-\,\,\frac{\partial\,  Z\left(Y  \right)}{
\omega_0\,\partial\,Y}\,\,=\,\,\,
Z^2\left(Y\right)\,\left(\,1\,\,-\,\,Z\left(Y \right)\right)\,.
\eeq
Let define $Z_0\,\equiv\,Z(Y\,=\,y_0\,=\,0)$.
\eq{TMNEQ} can be solved in the implicit form:
\beq \label{TMZSOL}
\frac{Z\,-\, 1}{Z}\,e^{ \,1/Z}\,\,=\,\,
\frac{Z_0 \,-\, 1}{Z_0}\,e^{\,1/Z_0}\,\,e^{\, \omega_0\,Y}
\eeq
It can be seen that $Z \,\rightarrow \,0$ and $N_A\,=\,1\,-\,Z\,\rightarrow\,1$
as $Y\,\rightarrow \,\infty$. 
In \fig{sol} we compare the amplitude $N_A$ calculated from the functional
of \eq{TMZSOL} with the amplitude obtained from \eq{SOLZ} the latter being
a solution to the BK equation of this simplified model.
The \fig{sol} shows a dramatic weakening  of the shadowing
due to the contribution
of the \fig{endi}-2 type diagrams. This effect starts to be noticeable
when the amplitude $N_A\,\simeq\, 0.5$. This point is usually associated with
the transition  to the saturation region. As rapidity increases the effect
strengthens and as a result, the amplitude approaches the black disc limit 
$N_A\,=\,1$ much slower compared to the standard behavior driven by the 
'fan' diagram of  \fig{endi}-1.

\section{Results and discussions}

In this paper we developed a new approach to high density QCD 
(Color Glass Condensate) based on a linear evolution equation
for QCD generating functional. We obtained three main results.

First, we derived \eq{LEQZF} which is a linear evolution equation for the 
generating functional   
summing all possible `fan' diagrams (see \fig{endi}-1).
This equation involves  functional derivatives with respect to initial
conditions which is the price paid for having a linear equation for 
non-linear dynamics. The linear equation for generating functional
can be rewritten in a non-linear form reproducing the very same
non-linear equation as derived by Mueller in Ref. \cite{MUUN}.

The second result of this paper is a generalization of \eq{LEQZF}. 
 \eq{NLEQZK} is a new linear equation which incorporates the 
Glauber-Mueller rescatterings and this way sums all possible diagrams of
the type shown in \fig{endi}-3.  \eq{NLEQZK} can be  reformulated as a 
new non-linear equation. The latter has a good  chance to be the 
correct non-linear equation describing the parton collective phenomena such as
 Color Glass Condensate.

From the  equations for the generating functional we were able to
obtain corresponding linear equations 
for the scattering amplitude. These equations involve functional derivatives
with respect to initial conditions. In principal, this should not be an
 obstacle since the
initial conditions are usually known and functional derivatives
can be replaced by ordinary derivatives. 
The striking advantage of the linear formulation that it allows us
to address a question of target correlations.
Namely those correlations which occur in the  interactions of  
`wee' partons with the target. 

It follows from our analysis that the linear equation can be reformulated
as standard non-linear evolution in two cases only. The first is when 
the interactions of `wee' partons are fully uncorrelated. When only the
'fan' diagrams are summed this results in the BK equation. The second case
includes target correlations but in a most simplistic way in which only
one correlation parameter is introduced independently of number of dipoles
participating in the interaction. We firmly believe 
that within the approach based on linear  equation for the 
scattering amplitude we can address a systematic 
analysis of target correlations and their influence on the 
value and energy dependence  of the scattering 
amplitude.

The third result of this paper is in the application of the ideas developed
above to interactions with realistic nuclei. We argued that it is not
correct to treat realistic (dilute) nuclei relying on  
the large $A$ approximation. In our work we relaxed this approximation
and instead introduced nucleus correlations in a model dependent way.
The proposed model is such simple  that the linear equation for the 
amplitude can be brought to a non-linear form. We obtained this way 
a new non-linear equation  (\eq{NLEQGFAA}) which is a generalization of the
BK equation for realistic nuclei. We hope that this new equation would lead to 
more reliable calculations  for ion-ion collisions and would
 influence  our understanding  of the RHIC data.

\section*{Acknowledgments}

We wish to thank  Asher Gotsman, Yura Kovchegov, Alex Kovner, 
Uri Maor and Urs Wiedemann for very fruitful discussions.

E.L.  is grateful to the DESY Theory Division for their
hospitality. E.L. is  indebted to the Alexander-von-Humboldt Foundation for the
award that gave him a possibility to work on low $x$ physics during the
last year.

This research was supported in part by the
GIF grant \# I-620-22.14/1999, and by the Israel Science Foundation,
founded by the Israeli Academy of Science and Humanities.

\end{document}